# Non-Ergodic-Induced Negative Differential Piezoresponse in Relaxor Ferroelectrics

Cecile Saguy[a], Benjamin Kowalski[b], Alp Sehirlioglu[c] and Yachin Ivry[a, d, e, *]

[a] *Solid State Institute, Technion−Israel Institute of Technology, Haifa 3200003, Israel*

[b] *NASA Glenn Research Center. Cleveland, Ohio 44135, USA*

[c] *Department of Material Science and Engineering, Case Western Reserve University, Cleveland, Ohio 44106, USA*

[d] *Department of Materials Science and Engineering and Solid State Institute, Technion−Israel Institute of Technology, Haifa 3200003, Israel.*

[e] *The Nancy and Stephen Grand Technion Energy Program (GTEP), Technion – Israel Institute of Technology, Haifa, 3200003, Israel*

\* Email: ivry@technion.ac.il

**Abstract**

Relaxor ferroelectrics exhibit a unique competition between long-range and short-range interactions that can be tuned electrically which prioritizes these materials in a broad range of electro-mechanical energy-conversion technologies, including biomedical imaging and electric-charge generators. Here, we demonstrate differential negative piezoresponse by utilizing the short-range interactions in relaxor ferroelectrics. The effect was observed over a broad temperature range with local piezoresponse spectroscopy in unpoled samples, while no negative piezoresponse was observed when the material was pre-scan poled. These measurements suggest that the effect, that is promising for power-generation applications, originates from non-ergodic behavior. Complementary macroscale impedance and dielectric constant measurements as a function of temperature and frequency supported the mesoscopic findings. Bearing in mind the direct relationship between piezoresponse and capacitance, relaxor ferroelectrics appear as an excellent platform for the emerging technology of low-power negative-capacitance transistors.

Much effort has been put recently[1,2] in utilizing the attractive effect of negative capacitance in low-power memory devices and computation technologies. Capacitance is charge (*Q*) storage upon voltage (*V*) increase: $C=dQ/dV$, which is the core of a broad range of technologies, including microelectronics, energy storage, signal processing and sensing. In negative capacitance, voltage and charge exhibit an opposite trend, so that if introduced in a common transistor, current increases faster than the relative gate voltage, increasing the subthreshold slope.[3] This counter-intuitive effect is equivalent to an elastic membrane in a pressurized water pipe that is stretched towards the high-pressure side. It has been proposed that in ferroelectrics, charge is decreasing upon polarization domain switching, giving rise to the desired charge accumulation near the gate. The exact fundamental negative-capacitance mechanism is still under debate.[4] It is largely believed, however, that the effect is possible locally, while the ferroelectric returns to exhibit normal capacitance at the larger scale.[5] It is assumed then that the main contribution in real negative capacitance stems from the fact that the polarization-induced surface charge in ferroelectrics ($\sigma_P$) is not always compensated locally.[6–9] For instance, charged domain walls, and chemical interactions at the ferroelectric surface and interface as well as at the domain walls impose a mismatch between the presumably surface compensating charge ($\sigma_s$) and the polarization-induced charge distribution. This modulation gives rise, in turn, to excess accumulated charge that is available near the capacitor electrode and is readily released.[7] Bearing this mechanism in mind, Íñiguez et al.[6] listed properties that are required for significant negative capacitance. A ferroelectric with small domains is essential, typically available in thin films, to allow substantial domain-wall charge accumulation as well as wall-wall interactions.[10] Smaller domains are preferable due to the relatively higher domain-wall volume. High domain-wall energy is also required. Sandwiched ferroelectrics or a multilayer stack are needed for surface charge accumulation. Lastly, a broad temperature range of a stable multidomain region is thermodynamically preferable.[11] To-date, realizing a system with the above requirements has remained a non-trivial task, especially when the technological relevance is considered. Ferroelectric polymers and perovskites are preferred due to their relatively well-known properties and thin-film preparation methods.[12] Nevertheless, polymer stability, perovskite metallization, and their chip integration are limited, hindering device compatibility. Thus, much focus has been given lately to fluorite ferroelectrics[13] which grow well on a Si chip at relatively low temperatures and are technologically attractive. Nevertheless, the non-conventional ferroelectric-related properties of these hafnia-based materials are not yet well understood, raising the need for other ferroelectric materials that demonstrate the desired negative-capacitance behavior. Herein, relaxor ferroelectrics are proposed as a potential, nearly ideal system for negative capacitance. Relaxor ferroelectrics comprise nano-domains or polar nano-regions (PNRs).[14] The PNRs exist at a broad temperature range. Although the polarization within each PNR is switchable relatively easily, these domains or quasi-domains are pinned with high effective domain-wall energy. Relaxor ferroelectrics maintain these properties also at the bulk geometry and may therefore expand the potential applications of negative capacitance beyond thin films. As opposed to conventional ferroelectrics that exhibit a balance between local and macroscopic interactions, the dominant interaction in relaxor ferroelectrics can be either short range or long range, depending on the material history. That is, if the relaxor ferroelectric is subject to high electric fields, the dipoles will arrange in a preferable orientation to form macroscopic polarization. Nevertheless, if the material is heated towards the paraelectric state without being subject to an external electric field, the polarization is then carried out locally, via the PNRs. The dependence of the macroscopic behavior on whether the material was first subject to temperature variation or to electric poling has been documented.[15] However, local characterization of this unique behavior has remained elusive. The system of interest is (1-

x)Pb(Mg$_{1/3}$Nb$_{2/3}$)O$_3$ - xPbTiO$_3$ (PMN-PT). The prototypical PbMg$_{1/3}$Nb$_{2/3}$O$_3$ (PMN), traditionally thought to comprise PNRs in a non-polar matrix but has been also discussed in terms of local competition between local antiferroelectric correlations and local ferroic order. Addition of Ti$^{4+}$ on the B-site, forming (1-x)Pb(Mg$_{1/3}$Nb$_{2/3}$)O$_3$ – xPbTiO$_3$ (PMN-PT), results in increased off-centering and thus polarizability, consequently breaking the pseudo-cubic symmetry and producing rhombohedral (R) symmetry (or reduction of antiferroelectric correlations).[16] The R phase is also resolved as a monoclinic phase especially near the morphotropic phase boundary (MPB), changing from space group Cm to Pm with increasing PT content until the MPB. In the present work, mesoscale characterization of the non-ergodic to ergodic behavior of bulk 0.72Pb(Mg$_{1/3}$Nb$_{2/3}$)O$_3$ – 0.28PbTiO$_3$ (0.72PMN-0.28PT) has been demonstrated. Correlated PFM and switching spectroscopy PFM (SSPFM) measurements show clear difference in temperature-dependent behavior of the local polarization distribution on whether the material was first subject to an external electric field. SSPFM hysteresis loops of the pre-scan unpoled relaxor demonstrate negative differential piezoresponse, and higher switching voltages. PFM imaging revealed that the negative piezoresponse is often accompanied by local coalescence of small domains. Contrariwise, when pre-scan poling was performed prior to the heating, the SSPFM performed on the poled regions resulted in a conventional ferroelectric hysteretic shape and the switched domain did not interact much with its surrounding. The PFM results were correlated with independent macroscopic impedance and dielectric constant measurements as a function of temperature and frequency.

## Results

Figure 1a shows the dielectric constant as a function of both temperature and frequency for an unpoled bulk crystal of PMN-PT. Similar to a relaxor, a broad phase transformation peak ($T_{max}$) is observed rather than a sharp Curie Temperature ($T_C$). However, the frequency dispersion is much smaller than in a prototypical PMN; here it is 4° with three decades of frequency increase, while it is 18° for pure PMN. The composition is near the R-side of the MPB phase boundary region where the R/Cm to T (rhombohedral/monoclinic to tetragonal) phase-transformation temperature and $T_{max}$ approach each other,[17] as seen in the PMN-PT phase diagram shown in Figure S1.[18,19] The shoulder in the lower temperature side can be attributed to this phase transformation or to the presence of freezing temperature ($T_f$) in a relaxor, below which the frequency dispersion ends and a non-ergodic (NE) relaxor phase forms. Poling results in macroscopic piezoelectricity and the associated resonance peak in impedance spectra. The poled state disappears at this temperature, corresponding to the shoulder resulting in depoling. This depoling state is presented in Figure 1b (dashed line) with the disappearance of the resonance peak in the phase-angle contour plot and is not accompanied by a significant jump in dielectric constant (Figure 1a). Note that the phase-angle contour plot is considered a good representation of the depoling behavior.[20]

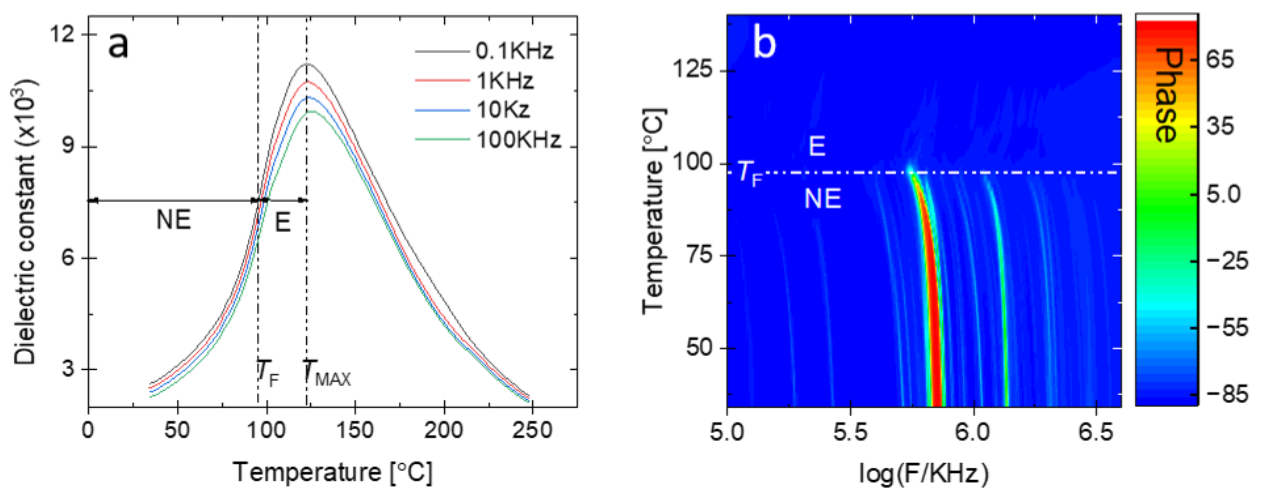

Figure 1. Macroscopic dependence of ergodicity on pre-poling in PMN-PT. (a) Dielectric constant as a function of temperature and frequency for an unpoled bulk crystal and (b) the phase angle as a function of temperature and frequency for a poled single crystal. $T_F$ and $T_{MAX}$ are the freezing and maximum dielectric constant temperature, respectively. NE and E denote non-ergodic and ergodic, respectively.

Once the ergodicity was characterized macroscopically, mesoscale characterization was done to examine the piezoresponse which is related directly to the capacitance behavior.[21] Variable-temperature piezoresponse hysteresis-loop measurements and domain imaging were performed with and without pre-scan poling to complement the macroscopic impedance measurements and examine the ergodicity in these materials at the mesoscopic scale. Although the frequency dispersion obtained from macroscopic measurements presented in Figure 1 is relatively small (4° in three decades of frequency), the mesoscopic scale characterization of the unpoled sample carried out by piezoresponse force microscopy (PFM) amplitude and phase signals show the coexistence of ferroelectric striped and curved domains as well as nano-domains characteristic of relaxors (Figure 2a).[22,23] Such differences in macroscopic and mesoscopic characterization is not uncommon.[24] Pre-scan poling was done via scanning with a conductive AFM tip in an area that is much larger than the unpoled domain size ('square in a square,' Figure 2b) at room temperature (see Methods for details). The polarization distribution after local hysteresis measurements was characterized in the downwards (inner square) pre-scan polarized region. When the SSPFM cycle was terminated by negative polarity, isolated domains with clear domain walls were formed in the inspected region (highlighted by a red arrow in Figure 2c). Figure 2d shows piezoresponse hysteresis loops that were obtained at SSPFM measurements, demonstrating a typical ferroelectric behavior. A complete temperature-dependent hysteresis loop set throughout the phase transition following the large-area pre-scan poling procedure is given in Figure S2. While the loops taken from room temperature to 100 °C show typical ferroelectric behavior, part of those taken at temperatures above 100 °C demonstrate a diamond shape, which is associated with relaxor behavior.[25]

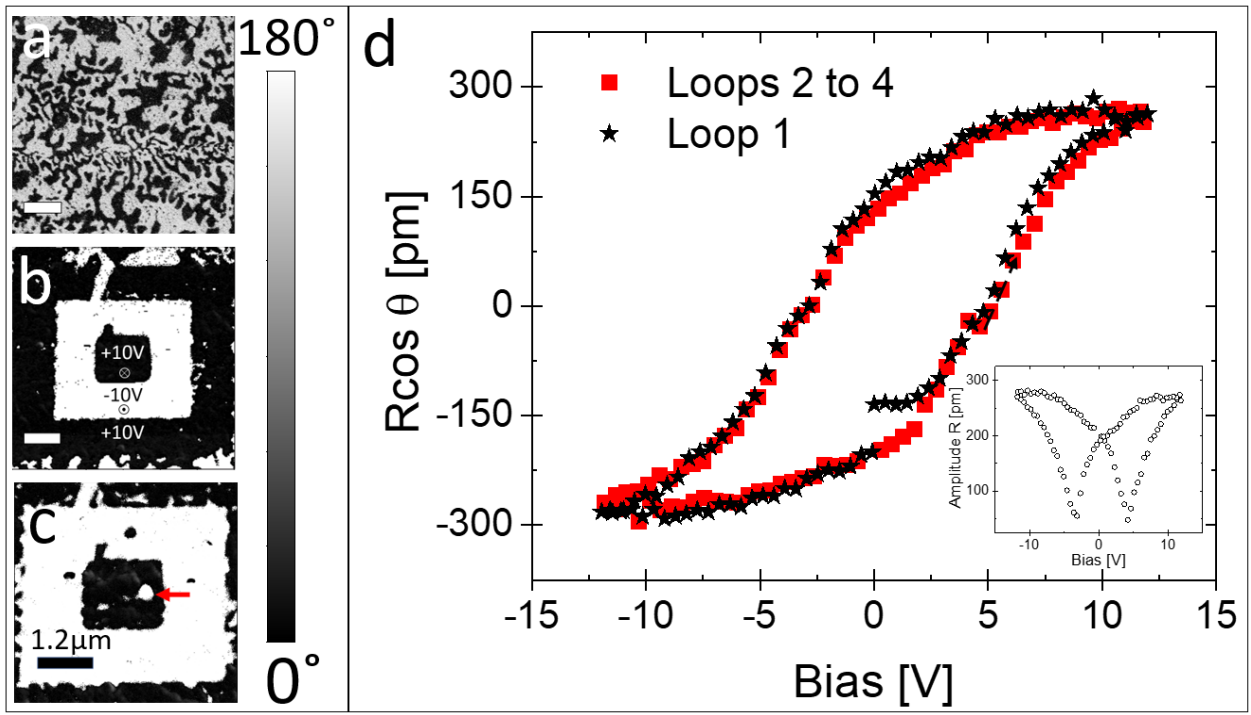

Figure 2. Polarization distribution and local hysteretic behavior in pre-scan poled PMN-PT. PFM phase of (a) an unpoled sample; (b) after poling a large 'square in a square' area and (c) zoom-in of the same place, after local SSPFM measurements. Scale bars of (a-c) are 1.2 µm. (d) Piezoresponse hysteresis loop, measured at the highlighted spot (red arrow) in (c). Insert: simultaneously measured amplitude signal hysteresis loop.

To examine the ergodicity in these materials at the mesoscopic scale, the experimental protocol was repeated. This time, however, the SSPFM was performed in an area that included unpoled domains only and was not pre-scan poled. Figures 3a and 3b show the domain distribution at the unpoled state and after local SSPFM was done, respectively. A careful look at the domain redistribution reveals that as opposed to local domain switching observed in the pre-scan poled sample, here, several domains merged together during the SSPFM measurement. Figure 3c shows that here, contrary to the smooth hysteresis loop obtained in the pre-scan poled sample (Figure 2d), the loop comprises a region in which the piezoresponse-voltage slope is negative, as confirmed by the derivative displayed around -2 V in the inset of Figure 3c. This negative piezoresponse arises from changes in the absolute value, i.e., the amplitude signal (Figure 3d), with only a negligible trace at the phase signal (Figure 3e). Moreover, the amplitude loop is asymmetric with respect to the remanent polarization, while the negative and positive coercive fields are rather similar. This asymmetry is clearly noticeable where transient strain-related states appear around 0 bias, a behavior that is most probably related to internal random fields causing a preferential polarization direction, which affects the piezoresponse.

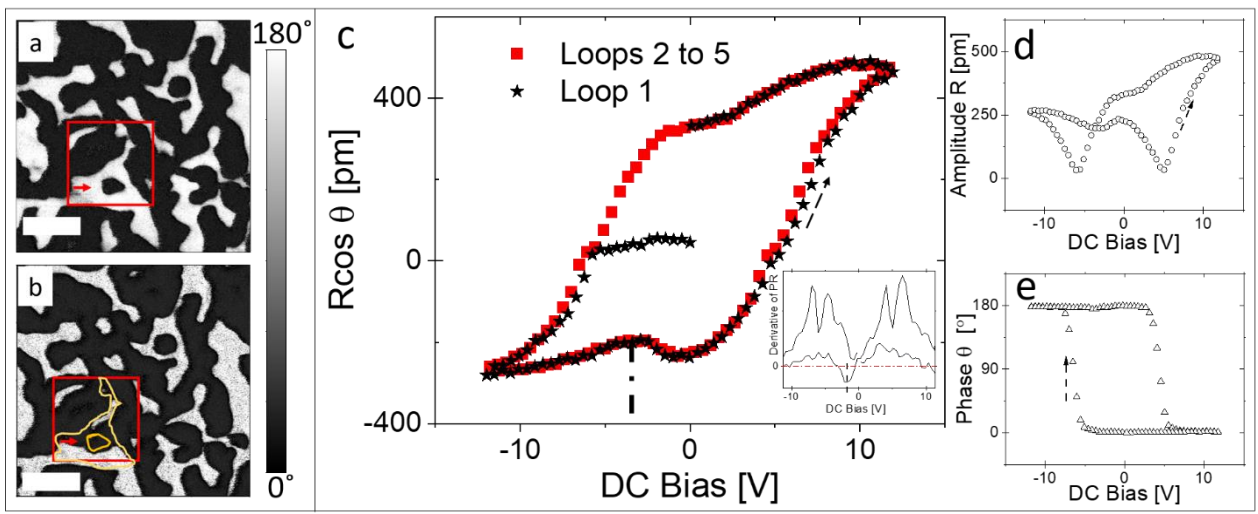

Figure 3. Polarization distribution and local hysteretic behavior in pre-scan unpoled PMN-PT. Polarization distribution (a) before and (b) after a local SSPFM measurement. Red arrows show the point where the hysteresis was done. The yellow lines indicate the contour of the original domains. The data scale on (a) and (b) is 0.5µm. Piezoresponse (c) mixed signal; (d) amplitude; and (e) phase hysteresis loop measured at the highlighted spot in (b). The dashed arrows show the direction of the loop, amplitude and phase. Dashed line indicates the bias at which negative piezoresponse is observed. Inset of (c) shows the derivative of the hysteresis loop, exhibiting differential negative piezoresponse. Data were collected at 60 °C.

The existence of a negative slope in the hysteresis loop was observed in the unpoled samples consistently while heating the sample also much above room temperature (Figure S4). The domain imaging accompanied to the SSPFM measurements show that the negative slope was often accompanied with domain merging also in the other measurements. Contrariwise, hysteresis loops obtained from pre-scan poled samples were smooth and with no negative slope during the similar temperature range. The switched domains formed during the spectroscopy characterization in the pre-scan poled samples were isolated with clear domain walls.

**Discussion**

The above results suggest in several independent ways that the origin of the negative differential piezoresponse stems from the high-density of domain walls. This is reminiscent of studies that relate negative capacitance to high domain-wall concentration, in which available uncompensated charge gives rise to misalignment and difference in redistribution time between polarization and screening charges.[6] One should bear in mind also the direct relationship between polarization and electromechanical coupling in ferroelectrics. First, negative piezoresponse was not observed in macroscopically pre-scan poled samples (Figure 2) or in common perovskite ferroelectrics where there are no small domains,[26] but only in samples that were not pre-scan poled (Figure 3). Secondly, the dependence on small domains was highlighted by the observed coupling between negative piezoresponse and domain merging. Here, when the SSPFM was performed closer to a domain with similar end polarization that is closer than 200 nm, both negative piezoresponse and domain merging were observed.

PMN-PT is typically considered a relaxor-based system for stoichiometries close to the MPB. Away from the MPB itself, as in the current system, even though there is not always a consensus about the relaxor characteristics as well as exact polymorphs forming, the high domain density and labyrinth structure as well as the domain traces in the topography are all typical for such PMN-PT compositions.[27] This was exemplified here in the dielectric measurements where the broad phase transformation was accompanied by only a small frequency dispersion and a temperature where this dispersion disappears below the $T_f$ (Figure 1). The transition from the relaxor to a ferroic behavior is not a sharp transition but a gradual change of structural and performance characteristics.[19,28] Thus, while macroscopic measurements may be controlled by dominant features, locally, the behavior can be different and show dependence to secondary features. Samples were heated close to the temperatures where frequency dispersion starts indicating either proximity to the ergodic state or partial phase transformation to tetragonal polymorph. In addition to the different switching domain pattern and hysteretic behavior between pre-scan poled and unpoled samples (Figures 2-3, S2-S4), the non-ergodic behavior is further supported by careful analysis of the PFM data. Figure S3 shows that upon heating, the long-range ferroelectric order of poled large-scale domains remains rather stable. Yet, these large domains do not remain homogeneous. Rather, nano-domains emerge in the pre-scan poled domains and their density increases with temperature. The so-formed nano-domains remain stable also upon cooling back to room temperature. This nano-domain appearance (Figure S3) which involves the formation of a larger number of isolated domain walls is different than the typical domain relaxation in ferroelectrics that follows a more homogeneous shrinkage or growth of the domains at the domain boundaries.[29] This observation of partial local short-range-interaction at high temperatures indicates a non-ergodic-to-ergodic transition.

The existence of a gradual non-ergodic-to-ergodic transition at high temperatures is supported not only from PFM imaging, but also from the piezoresponse spectroscopy measurements. Below 100 °C, typical ferroelectric hysteretic behavior is observed in pre-scan poled areas (Figure S2). Above 100 °C, a gradual transition is observed, in the sense that some points exhibit typical relaxor diamond-shape hysteresis loops, while well-defined ferroelectric shape is measured at other points.[25] This behavior is different from the abrupt vanish of an hysteretic piezoresponse in the case of a structural phase transition, *e.g.*, from rhombohedral to tetragonal (or tetragonal to cubic). This spatial inhomogeneity of the hysteretic behavior is thus related to the fact that partial and local depoling of the pre-scan poled area has occurred with increasing the temperature above 100 °C, as shown in Figure S3. This conclusion is supported also by the hysteretic behavior of unpoled areas, which demonstrated spatial inhomogeneity also at lower temperatures. That is, in unpoled areas, while some hysteresis loops exhibit negative differential piezoresponse even upon heating to high temperatures, loops taken on other location display the diamond shape specific to relaxor or the well-defined ferroelectric behavior (Figure S4). These findings are consistent with previous literature that indicates that in relaxors, the mechanism of tip bias induced phenomena depends largely on its interaction with defects and the relation between the effective diameter of the tip and the scale of characteristic inhomogeneity in the material.[30] Likewise, the hysteresis loops indicate the existence of domain pinning or movable domains nearby the tip and random fields, which are typical for highly disordered ferroelectrics and relaxors.[25]

Figure S5 shows that the coercive field of the unpoled sample is consistently higher than that of the pre-scan poled sample, indicative of a higher built-in field that strongly pins local domain nucleation and growth in the unpoled sample. Moreover, Figure S5 shows that upon heating, the coercive field increases up to 90 °C and then remains stable for a broad temperature range until the phase transition is completed. This behavior indicates that the unpoled domains become more stable with increasing temperature, an inherent drive to behave as an ergodic material. As the temperature is lowered away from the freezing temperature, there is a higher drive to form longer range interactions. Thus, the change in non-ergodic to ergodic characteristics is a continuous one and once the ergodic phase forms, the coercive field remains rather constant.

**Conclusions.**

Negative differential piezoresponse was demonstrated in unpoled PMN-PT, using the short-range-interaction that exists in relaxor ferroelectric. SSPFM hysteresis loops taken randomly over a wide range of temperatures in the unpoled sample show negative differential piezoresponse at a peculiar bias. This phenomenon that is attributed to the non-ergodic behavior of the material was found to be reproducible at increasing temperatures. The occurrence of negative differential piezoresponse is often accompanied by merging of nano-domains located in the vicinity of an AFM tip, pointing out that the origin of the negative differential piezoresponse stems from the high-density of domain walls. When pre-scan poling is applied to the sample at sufficiently high bias, long-range ferroelectric order is demonstrated by means of domain distribution. The hysteresis loops taken on the negatively- and positively-poled area display well-defined ferroelectric behavior with clear onset for nucleation. The pre-scan poled regions are globally stable with increasing temperature. However above 100 °C, partial local depoling takes place, yielding to a relaxor-like diamond-shape hysteresis loops in addition to the common ferroelectric hysteresis loop behavior, demonstrating a non-ergodic-to-ergodic transition. Given the relationship between piezoresponse and polarization as well as between polarization and capacitance,[31] the negative differential piezoresponse detected here in relaxor ferroelectrics renders an excellent platform for the emerging technology of low-power negative-capacitance transistors. The significance to the data technology becomes even more pronounced when bearing in mind that the rich domain dynamics that affects the negative piezoresponse is dominant at the mesoscopic scale, which is in turn is technologically relevant.

**Methods.**

$0.72Pb(Mg_{1/3}Nb_{2/3})O_3$-$0.28$ $PbTiO_3$ single crystals were obtained from CTS Corporation where they were grown using a modified Bridgman technique.

Impedance and dielectric constant measurements were carried out by using an Agilent 4294A impedance analyzer. A modified Agilent 16334A sample connection was used to conduct measurements as a function of temperature in a modified Carbolyte furnace. Data were collected via a custom program using LABVIEW® (National Instruments Corp., Austin, TX) software.

PFM imaging and spectroscopy were done with DART PFM on an MFP-3D Infinity AFM system (Asylum Ltd by Oxford Instruments) at a drive amplitude of 1V. The AFM tips used were conductive PtSi-FM from Nanosensors Ltd. with a resonance frequency around 70 kHz and a force constant of 2.5N/m. Pre-poling was achieved by applying a +/-10 V bias between a conductive tip (bias applied to the tip) and a continuous bottom electrode so that a positive bias induces a downwards polarization area whereas a negative bias generates an upwards polarized region. Here, a 5x5 $\mu m^2$ (256 lines including each one 256 points) square was first written at +10 V. Then a smaller square of 3x3 $\mu m^2$ (256 lines including each one 256 points) was written at -10 V at the same frequency, and at last a square of 1x1 $\mu m^2$ (128 lines including each one 128 points) was written at +10 V. All the squares were written at a frequency of 1 Hz. The SSPFM hysteresis loops were measured at a frequency of 0.2 Hz and the results of 4 to 6 cycles were averaged.[32,33]

**Supporting Information**

Phase diagram of PMN-xPT system correlated with dielectric measurements showing that the shoulder in the lower temperature side of the dielectric constant can be attributed to rhombohedral/monoclinic to tetragonal phase transformation or to the presence of freezing temperature ($T_f$) in a relaxor, below which the frequency dispersion ends and a non-ergodic (NE) relaxor phase forms (Figure S1). Local hysteretic behavior in pre-poled (Figure S2) and unpoled PMN-PT (Figure S4) as a function of temperature. Piezoresponse force microscopy images showing the partial local depoling process following gradual heating to high temperature (Figure S3). Coercive bias as a function of temperature for pre-poled and native (unpoled) PMN-PT (Figure S5) (PDF).

**Acknowledgments**

This work was made possible through United States - Israel Binational Science Foundation (BSF) Grant no: 2020325. The Technion team acknowledges support also from the Zuckerman STEM Leadership Program and the Grand Technion Energy Program. We also thank Dr. Hemaprabha Elangoven for fruitful discussions.

**Supplementary Information**

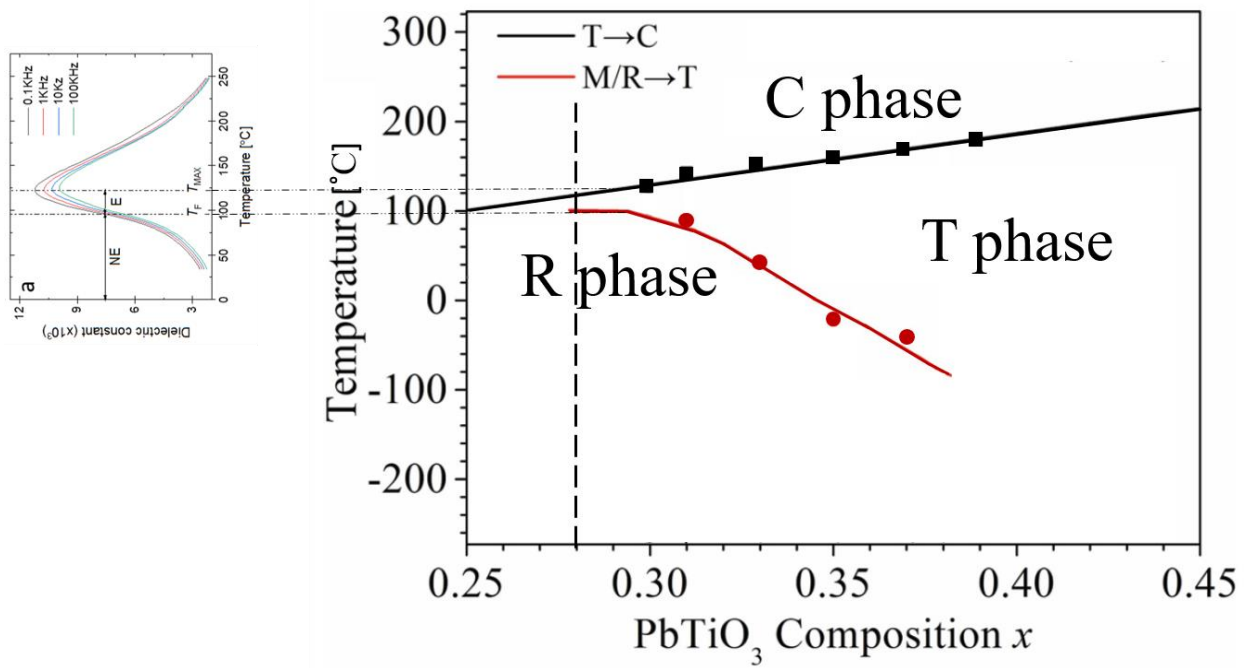

**Figure S1**. **Phase diagram of PMN-xPT system correlated with dielectric measurements**. The dashed black line indicates the composition of the PMN-PT (x=0.28) sample used in this paper. The composition is near the R-side of the MPB phase boundary region where the R/Cm→T (rhombohedral/monoclinic to tetragonal) phase-transformation temperature and $T_{max}$ approach each other (Noheda et al. Phys. Rev. B 66, 054104 (2002)). The shoulder in the lower temperature side of the dielectric constant can therefore be attributed to this phase transformation or to the presence of freezing temperature ($T_f$) in a relaxor, below which the frequency dispersion ends and a non-ergodic (NE) relaxor phase forms.

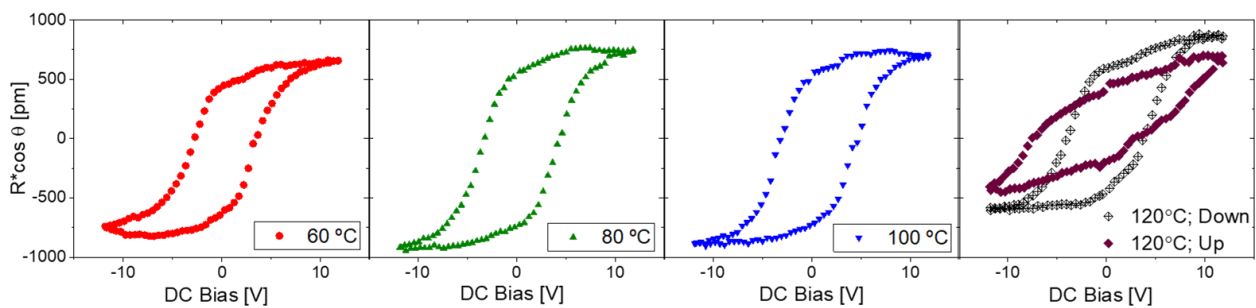

**Figure S2**. **Local hysteretic behavior in pre-poled PMN-PT as a function of temperature.** The mixed signal shows typical ferroelectric behavior till 100 ˚C with clear nucleation onset followed by an abrupt increase or decrease of piezoresponse. Above this temperature the hysteresis behavior is location dependent, resembling that of a relaxor or a common ferroelectric.

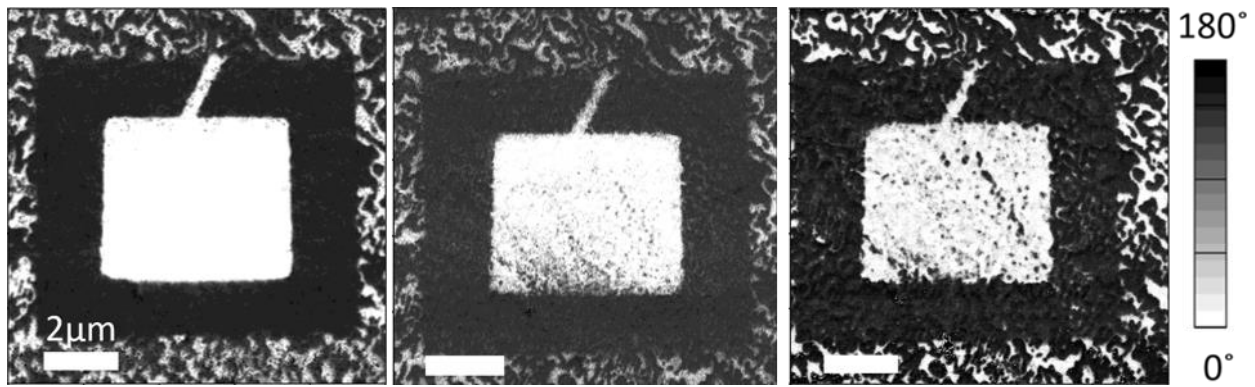

**Figure S3. Partial local depoling following gradual heating to high temperature**. Polarization distribution measured by the PFM phase signal, (a) after poling at room temperature; (b) after heating to 80 ˚C and (c) followed heating to 120 ˚C and cooling back to room temperature. The general structure of pre-scan poled large-scale domains remain rather stable, yet, not homogeneous. Note the appearance of downwards polarized nano-domains (black spots) within the negative polarized (upwards) pre-scan poled area as a result of the heating. This nano-domain distribution within the pre-poled area is different than the typical domain relaxation in ferroelectrics that follows a more homogeneous shrinkage or growth of the domains at the domain boundaries. The presence of nano-domains with preferential downward polarization confirms the downwards polarization stability as compared to upwards polarization.

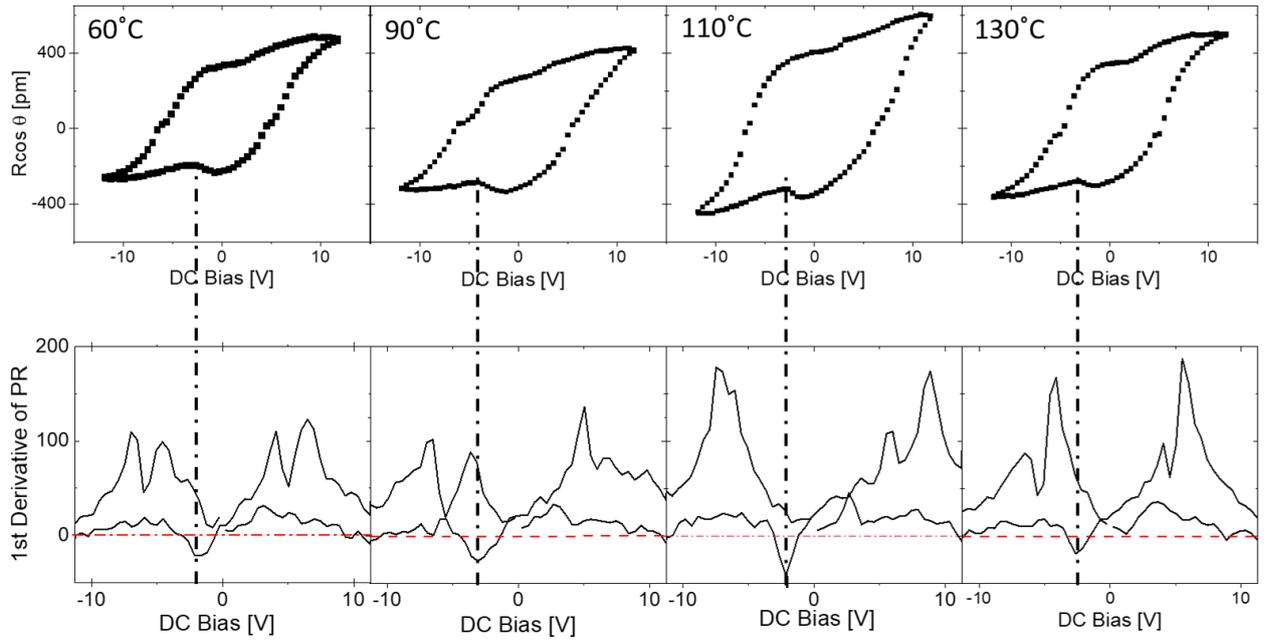

**Figure S4. Local hysteretic behavior in native PMN-PT as a function of temperature.** The negative differential piezoresponse in the mixed signal exists at increasing temperatures. It is marked by the black dashed dot lines in the first derivative of the piezoresponse (PR) and in the hysteresis loops.

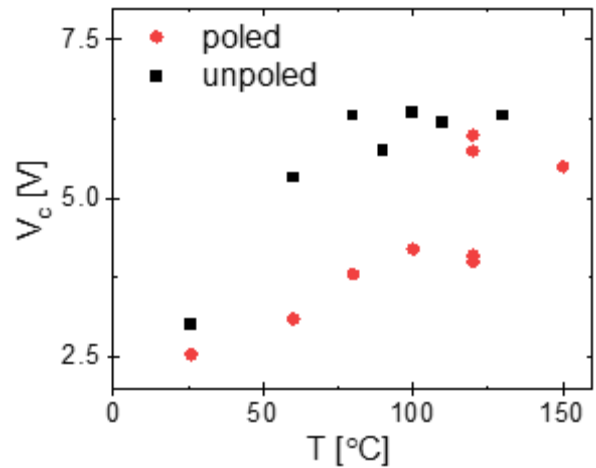

**Figure S5. Coercive bias as a function of temperature for pre-poled and native (unpoled) PMN-PT.** In native PMN-PT, the coercive bias saturates above 90 °C. In poled samples, the coercive field above 100 °C is location dependent, the high values (around 6V) corresponding

to a piezoresponse response characteristic of relaxor or unpoled sample and the lower saturated values (4V) to a ferroelectric piezoresponse behavior.